\documentclass{aastex}
\usepackage{emulateapj5}
\usepackage{apjfonts}
\usepackage{epsf}
\usepackage{amsmath}

\def\ce{\ifmmode {\cal E} \else ${\cal E}$\ \fi}

\def\cv{\ifmmode {\cal V} \else ${\cal V}$\ \fi}

\def\dotmb{\dot{M}_{\rm Bondi}}
\def\dotmedd{\dot{M}_{\rm Edd}}
\def\pe{\langle\epsilon\rangle}

\def\sunm{M_{\odot}}
\newcommand{\kms}{\ifmmode {\rm km\ s}^{-1} \else km s$^{-1}$\ \fi}
\newcommand{\ergs}{\ifmmode {\rm erg\ s}^{-1} \else erg s$^{-1}$\ \fi}
\newcommand{\tes}{\ifmmode \tau_{\rm es} \else $\tau_{\rm es}$\ \fi}
\newcommand{\tk}{\ifmmode \tau_{\rm K} \else $\tau_{\rm K}$\ \fi}
\newcommand{\vfwhm}{\ifmmode V_{\mbox{\tiny FWHM}} \else
            $V_{\mbox{\tiny FWHM}}$\fi}
\newcommand{\msun}{\ifmmode M_{\odot} \else $M_{\odot}$\ \fi}
\newcommand{\afe}{\ifmmode {\mathcal A_{\rm Fe}} \else${\mathcal A_{\rm Fe}}$\ \fi}

\newcommand{\etaa}{\ifmmode \eta_a \else $\eta_a$\ \fi}
\newcommand{\etaf}{\ifmmode \eta_{{_f}} \else $\eta_{{_f}}$\ \fi}
\newcommand{\ledd}{\ifmmode L_{\rm Edd} \else $L_{\rm Edd}$\ \fi}
\newcommand{\lx}{\ifmmode L_{\rm 2-10keV} \else  $L_{\rm 2-10keV}$\ \fi}
\newcommand{\hb}{\ifmmode H\beta \else H$\beta$\ \fi}
\newcommand{\mbh}{\ifmmode M_{\rm BH}  \else $M_{\rm BH}$\ \fi}

\journalinfo{The Astrophysical Journal Letters}

\begin{document}

\title{Feedback Limits Rapid Growth of Seed Black Holes at High Redshift}

\author{Jian-Min Wang\altaffilmark{1}, Yan-Mei Chen\altaffilmark{1} and Chen Hu\altaffilmark{2,1}}

\altaffiltext{1}{Key Laboratory for Particle Astrophysics, Institute of
High Energy Physics, Chinese Academy of Sciences, Beijing 100039, China.
wangjm@mail.ihep.ac.cn} 
        
\altaffiltext{2}{National Astronomical Observatories of China, Chinese Academy of Sciences,
Beijing 100012, China}

\slugcomment{to appear in The Astrophysical Journal Letters}
\shorttitle{Feedback of the Seed Back Hole Growth}
\shortauthors{WANG, CHEN \& HU}

\begin{abstract}
Seed black holes formed in the collapse of population III stars have been invoked to explain the presence of 
supermassive black holes at high redshift. It has been suggested that a seed black hole can grow up to 
$10^{5\sim 6}\sunm$ through highly super-Eddington accretion for a 
period of $\sim 10^{6\sim 7}$ yr between redshift $z=20\sim 24$. We studied the feedback of 
radiation pressure, Compton heating and outflow during the seed black hole growth. It is found that its
surrounding medium fueled to the seed hole is greatly heated by Compton heating.
For a super-critical accretion onto a $10^3\sunm$ seed hole, a Compton sphere (with a temperature 
$\sim 10^6$K) forms in a timescale of $1.6\times 10^3$yr so that the hole is only supplied by a 
rate of $10^{-3}$ Eddington limit from the Compton sphere.  Beyond the Compton sphere, the kinetic feedback 
of the strong outflow heats the medium at large distance, this leads to a 
dramatical decrease of  the outer Bondi accretion onto the black hole and avoid the accumulation of the 
matter. The highly super-critical accretion will be rapidly halted by the strong feedback.
The seed black holes hardly grow up at the very early universe unless the strong feedback can be avoided. 
\end{abstract}

\section{Introduction}
Black holes are regarded as an extremely important population in modern cosmological physics.
The reionization of the Universe may get started from $z\sim 17$ deduced from {\em Wilkinson Microwave
Anisotropic Probe} ({\em WMAP})  which gives Thomson scattering depth $\tau=0.17\pm 0.04$ (Spergel et al. 
2003). Such an early reionization epoch needs a large population 
of seed black holes collapsed from population III stars (Madau et a. 2004). Second, the discovery of the
currently known highest redshift quasar, SDSS 1148+3251 at $z=6.4$ (roughly 1 Gyr) 
from {\em Sloan Digital Sky Survey} (SDSS; Fan et al. 2001) indicates 
that there are already supermassive black holes with $\mbh>10^9\sunm$ (Netzer 2003,
Barth et al. 2003, Willott et al. 2003). What is the relation between the seed and supermassive black holes?
How to form supermassive black holes?

Rees' diagram shows several possible ways to form supermassive black holes (Rees 1984). A direct collapse of 
primordial clouds could form supermassive black holes after cosmic background radiation photons remove enough 
angular momentum through Compton drag (Loeb 1993, Loeb \& Rassio 1994). This scenario is favored by 
Ly$\alpha$ fuzz of the extended emission in quasar where the supermassive black hole has been formed and the
galaxy is assembling (Weidinger, Moller \& Fynbo 2004), especially the recent discovery of an isolated black 
hole of the quasar HE 0450-2958 without a massive host galaxy (Magain et al. 2005). Second,
a rapid growth of a seed black hole with highly super-Eddington accretion rates is used to explain the
existence of the black hole $>10^9\sunm$ at high redshift.  Third a compact cluster of main sequence stars, 
or neutron stars/black holes will inevitably evolve into a supermassive black hole (Duncan \& Shapiro 
1983, Quinlan \& Shapiro 1990), and this gets supports from the quasar's metallicity properties (Wang 2001).
The different ways to form a supermassive black hole may apply to different redshifts or environments. 

A rapid growth of seed black holes is quite a promising model to issue the formation of supermassive black holes 
at redshift $z\sim 6$ (Volonteri \& Rees 2005). The Bondi accretion rate is $\dot{m}=\dotmb/\dotmedd\sim 40$,
where $\dot{M}_{\rm Edd}=L_{\rm Edd}/c^2$ and $L_{\rm Edd}$ is the Eddington limit,  for 
a $10^3\sunm$ black hole surrounded by medium cooled by the hydrogen atomic lines. The 
seed black hole is able to grow up exponentially $\mbh=M_0\exp(\dot{m} t/t_{\rm Salp})$, where the Salpeter 
timescale $t_{\rm Salp}=0.45$Gyr. However, such a high accretion rate inevitably gives rise to strong interactions 
of radiation and outflows with the Bondi accretion flow. Consequently, the strong feedback seriously constrains 
the matter supply to the black hole, even stops the accretion.

In this Letter we discuss how the feedback impacts on the growth of the seed black holes. We find they
can not grow up between redshift $z=20\sim 24$ through accretion. The implications of the present results are 
discussed.

\section{Growth: Feedback Limit}

\subsection{Angular momentum and accretion onto seed black holes}
The very first population III stars rapidly evolve into intermediate mass black holes (IMBHs), $20<\mbh/\sunm<70$
and $130<\mbh/\sunm<600$ (Fryer et al. 2001, Omukai \& Palla 2003). These IMBHs, as shown in Volonteri \& Rees 
(2005), can rapidly grow up to supermassive black holes via Bondi accretion with a super-critical rate.

The strong tidal torque due to nearby clouds efficiently offer angular momentum $J$ to each other
(Peebles 1969, Barnes \& Efstathiou 1987). For a rigid virialized sphere with mass $M$, total
energy $E$ and radius $R$, its angular momentum is characterized by $J_0=GM^{5/2}/|E|^{1/2}$, where $G$ is the 
gravitational constant (Peebles 1969). It is useful to define a dimensionless parameter, 
$\lambda\equiv J/J_0=J|E|^{1/2}G^{-1}M^{-5/2}$, for a cloud with mass $M$, angular momentum $J$ and the total
energy  $E$. Following Oh \& Haiman (2002), a cold fat disk is formed with an isothermal exponential radial
distribution
\begin{equation}
n(R,h)=n_0\exp{\left(-\frac{2R}{R_d}\right)} 
         \sec^2\left(\frac{h}{\sqrt{2}H_0}\right),
\end{equation}
where $n_0$ and $H_0$ are the central density and vertical scale height of the disk at radius $R$, $R_d$ is 
the radial scale length. The central

\figurenum{1}
\centerline{\includegraphics[angle=-90,width=7.5cm]{fig1.ps}}
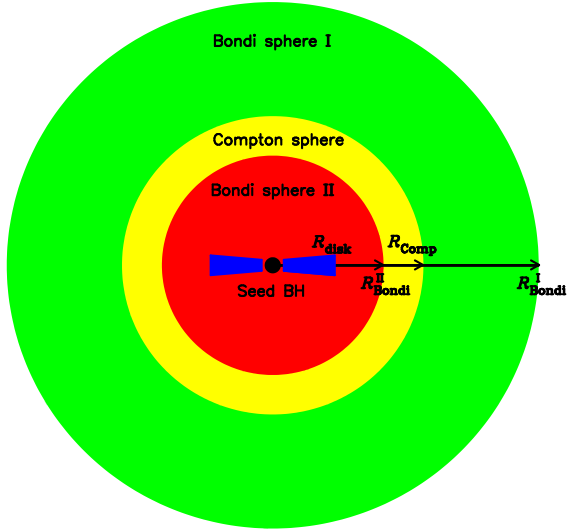
\figcaption{\footnotesize The scenario of a growing seed black hole. A tiny accretion disk is formed since the
accreting matter has angular momentum. The feedback from the tiny disk has strong influences on its
surrounding, leading to shrink the Bondi accretion radius and dramatic decreases of Bondi accretion rate
since the sound speed increases. The outflow developed 
from the disk has strong influences on the region beyond the Bondi sphere I. The super-Eddington accretion
is suddenly suppressed and the growth of the seed black hole stops (see detail in the text).}
\label{fig1}
\vglue 0.5cm

\noindent density is given by
\begin{equation}
n_0\approx 6\times 10^4f_{d,0.5}^2~\lambda_{0.05}^{-4}\left(T_3/8\right)^{-1}
             R_{\rm vir,6}^{-4}M_{\rm H,9}^2~~~{\rm cm^{-3}},
\end{equation}
where $f_{d,0.5}=f_d/0.5$ is the gas fraction settled into the isothermal disk, $\lambda_{0.05}=\lambda/0.05$, 
$T_3=T/10^3{\rm K}$, $R_{\rm vir,6}=R_{\rm vir}/6{\rm kpc}$ and 
$M_{\rm H,9}=M_{\rm H}/10^9\sunm$ are the virialized radius and mass of the halo, respectively. The disk's
radius scale is $R_d\sim \lambda R_{\rm vir}/\sqrt{2}$.

The Bondi accretion radius of the seed black hole is
\begin{equation}
R_{\rm Bondi}^{\rm I}=1.4\times 10^9R_{\rm G} ~m_3\left(T_3/8\right)^{-1},
\end{equation}
where $m_3=\mbh/10^3\sunm$, $R_{\rm G}=G\mbh/c^2$ and $c$ is the light speed. 
$R_{\rm Bondi}^{\rm I}$ is much smaller than the 
vertical scale of the cold fat disk. This means that the density within Bondi accretion radius is almost 
constant with a quasi-spherical geometry. As argued by Volonteri \& Rees (2005), the transverse velocity at 
the Bondi radius ($R_{\rm Bondi}^{\rm I}$)
is of the order of $\left(R_{\rm Bondi}^{\rm I}/R_d\right)v_{\rm D}$, where 
$v_{\rm D}$ is the rotation velocity of the fat gas disk at $R_d$.
Due to the specific angular momentum of the quasi-Bondi flow, a tiny disk 
will form within  $R_{\rm Bondi}^{\rm I}$. Figure 1 shows a cartoon of the present case with feedback. 
Assuming conservation of the specific angular momentum within $R_{\rm Bondi}^{\rm I}$,
we have the outer radius of the tiny accretion disk
\begin{equation}
R_{\rm disk}=6\times 10^2R_{\rm G}~ v_{\rm D,10}^2m_3^2\lambda_{0.05}^{-2}R_{\rm vir,6}^{-2},
\end{equation}
where $v_{10}=v/10\kms$, which is much smaller than the Bondi accretion radius.
The quasi-sphere flow will switch on 
the tiny disk with a Bondi accretion rate, given by  
$\dot{M}_{\rm Bondi}=\gamma 4\pi nm_p\left(G\mbh\right)^2/c_s^3$ (Bondi 1952)
\begin{equation}
\frac{\dot{M}_{\rm Bondi}^{\rm I}}{\dot{M}_{\rm Edd}}=188~\gamma m_3(n_4/4)(T_3/8)^{-3/2},
\end{equation}
where $c_s$ is the sound speed, $m_p$ is the proton mass, $\gamma$ is a dimensionless constant of order 
unity and $n_4=n/10^4{\rm cm^{-3}}$. The tiny accretion disk with such a high rate is characterized by  
photon-trapping self-similar structure within trapping radius, where the photon escaping timescale from 
the disk is equal to that of the radial moving to the black hole (Begelman \& Meier 1982, Wang \& Zhou 1999,
Ohsuga et al. 2003). We note that the disk radius ($R_{\rm disk}$) is comparable to the photon 
trapping radius $R_{\rm tr}\approx 3.2\times 10^2 R_{\rm G}\left(\dot{m}/180\right)$ 
(Wang \& Zhou 1999), indicating that the entire tiny disk may be advection-dominated via photon trapping 
process. According to the self-similar solution of the optically thick advection-dominated accretion flow 
(Wang \& Zhou 1999), the radiated luminosity is only a small fraction of the released gravitational energy 
via viscosity dissipation
\begin{equation}
L_{\rm Bol}\approx 4.0\times 10^{40}m_3\left[1+\frac{1}{4}\ln \left(\frac{\dot{m}}{180}\right)\right]~~~\ergs,
\end{equation}
where we have used the photon trapping radius $R_{\rm tr}$ implied above.
This is a saturated luminosity weakly dependent on the accretion rate. 
The total released luminosity is given by 
$L_{\rm grav}=1.4\times 10^{42}(\eta/0.1)\left(\dot{m}/10^2\right)m_3$ \ergs, most of them
are advected into the black hole.

The emergent spectrum from the slim disk is characterized by a universe shape of $F_{\nu}\propto \nu^{-1}$ 
(Wang \& Zhou 1999, Wang et al. 1999, see also Shimura \& Manmoto 2003, Kawaguchi 2003), the hard X-ray 
emission is highly dependent on the factor $f$, which indicates how many energy is released in the hot corona
(Wang \& Netzer 2003, Chen \& Wang 2004, hereafter, Socrates \& Davis 2005). Even $f=1\%$, the released energy 
from hot corona is close to Eddington limit. The physics related with the factor $f$ remains open. 
The Compton temperature reads $T_{\rm C}=\pe/4k$, where $k$ is the Boltzmann constant and $\pe$ is the
mean energy of photons from the disk which depend on $F_{\nu}$. The mean energy of the photons are in a 
range of $0.4\sim 10$keV for $f=0.01\sim 0.025$ (Wang \& Netzer 2003). We take the lower value $\pe=0.4$keV, 
so the medium will be heated up to $T_C\approx 10^6$K at least. 
It should be pointed out that the higher the Compton temperature 
is, the stronger the feedback from the tiny accretion disk has influence on its surroundings.

\subsection{Feedback: radiation pressure and Compton heating}
The radiation pressure acting on the medium inside the Bondi accretion sphere is of order 
$P_{\rm rad}\sim \tau_{\rm es}L/4\pi R^2c$, where the Thomson scattering depth is
$\tau_{\rm es}=Rn\sigma$ and $\sigma$ is Thomson cross section.
We find the ratio of $P_{\rm rad}/P_{\rm gas}=L\sigma/4\pi R ckT$, beyond the radius
$R_R=8.5\times 10^{7}R_{\rm G}~L_{40}T_4^{-1}m_3^{-1}$,
the radiation pressure is negligible, where $T_4=T/10^4{\rm K}$. 
Since the photon trapping effects, the radiated luminosity from the 
highly super-critical accretion disk is lower than Eddington limit by a factor of 3 (see equation 6). 
So radiation pressure hardly prevents the infalling matter from accretion onto
the black hole. The feedback from the radiation pressure is negligible.

The Bondi flow will be heated by the radiation from the tiny accretion disk. In turn, the heated medium
determines a new Bondi accretion rate switched on the tiny disk.
The thermal status of the photoionized medium can be conveniently described by 
the ionization parameter defined as $\Xi=L/4\pi R^2c nkT$ (Krolik, McKee \& Tarter 1981). We focus on a 
region with $\Xi\ge \Xi_{\rm C}=1.1\times 10^{3}T_{C,6}^{-3/2}$ (for pure hydrogen gas), 
here $T_{C,6}=T_C/10^6{\rm K}$, where the entire medium will be heated up to Compton temperature $T_C$.
It follows from
\begin{equation}
R_{\rm Comp}^{\Xi}=\left(\frac{L}{4\pi c\Xi_c nkT}\right)^{1/2}=2.8\times 10^{8}R_{\rm G}~L_{40}^{1/2}
    n_4^{-1/2}T_4^{-1/2}m_3^{-1},
\end{equation}
which is called the Compton radius (and sphere). It needs a Compton timescale 
$t_C=6\pi m_ec^2R_{\rm Comp}^{{\Xi}^2}/\sigma L\sim 1.3\times 10^{5}{\rm yr}$ to 
form such a Compton sphere.  However the Compton sphere is also constrained by the balance 
of the heating and infalling timescale of the Bondi flow. It follows from $t_C=t_{\rm infall}$, 
\begin{equation}
R_{\rm Comp}^{\rm infall}=\frac{\sigma L}{6\pi m_ec^2\beta_s c_s}
                         =3.1\times 10^7 R_{\rm G}~\beta_{s,0.1}^{-1}L_{40}T_4^{-1/2}m_3^{-1}
\end{equation}
where $\beta_s=v_R/c_s$ is the infalling velocity of the Bondi flow normalized by the sound speed and 
$\beta_{s,0.1}=\beta_s/0.1$. The effects of the infalling on the Compton sphere can be neglected if the
infalling velocity is less than $\beta_s \le 10^{-2}$. For $\beta_s=0.1$, we find 
$R_{\rm Comp}^{\Xi}>R_{\rm Comp}^{\rm infall}$. This means that {\em not } entire region with $\Xi>\Xi_c$ 
can be heated up to the Compton temperature ($T_C$). So the Compton radius is given by 
$R_{\rm Comp}=\min\left(R_{\rm Comp}^{\Xi},R_{\rm Comp}^{\rm infall}\right)=R_{\rm Comp}^{\rm infall}$, 
corresponding to a timescale of $1.6\times 10^3$yr. We note that 
$R_{\rm Comp}< R_{\rm Bondi}^{\rm I}\ll R_d\approx 0.2\lambda_{0.05}R_{\rm vir,6}$ kpc.

Within the Compton sphere, the seed black hole has a new Bondi accretion radius 
\begin{equation}
R_{\rm Bondi}^{\rm II}=\frac{G\mbh}{c_s^2}=1.1\times 10^7R_{\rm G}~T_6^{-1},
\end{equation}
and a Bondi accretion rate,
\begin{equation}
\frac{\dot{M}_{\rm Bondi}^{\rm II}}{\dot{M}_{\rm Edd}}=0.13~m_3 (n_4/4)T_6^{-3/2}.
\end{equation}
We note that this radius is still much larger than the tiny disk $R_{\rm disk}$ and 
this accretion rate is much smaller than
that in the pre-feedbacked. The timescale of the Compton heating at $R_{\rm Bondi}^{\rm II}$
is $t_C=2.0\times 10^2$yr.  That is to say the beginning Bondi accretion will be halted suddenly. After the
termination of the super-critical accretion, the hot plasma in Bondi sphere II will be cooled at a timescale of 
$1.8\times 10^4n_{C,2}^{-1}T_{C,6}^{1/2}$yr via free-free emission, where $n_{C,2}=n_C/10^2{\rm cm^{-3}}$ and
$T_{C,6}=T_C/10^6{\rm K}$. Therefore it is expected that there is a steady accretion onto the seed black hole, 
but with a sub-Eddington rate, which is self-regulated. 

When the Compton sphere forms, the pressure balance with its surroundings should hold, we have
$n_CkT_C=n_0kT_0$, where the subscript $"C"$ and $"0"$ represent Compton sphere and its pre-feedbacked 
surroundings, respectively. So we can get the Bondi accretion rate $\dot{M}_{\rm Bondi}^{\rm II}$ inside the 
Compton sphere in the form of $n_0$ and $T_0$,
\begin{equation}
\frac{\dot{M}_{\rm Bondi}^{\rm II}}{\dot{M}_{\rm Edd}}=1.3\times 10^{-3}m_3(n_{0,4}/4)T_{0,4}^{-3/2},
\end{equation}
where $n_{0,4}=n_0/10^4{\rm cm^{-3}}$ and $T_{0,4}=T_0/10^4{\rm K}$.
With such a low accretion rate, the seed black hole can not grow significantly.

Since $\dot{M}_{\rm Bondi}^{\rm I}\gg \dot{M}_{\rm Bondi}^{\rm II}$,
the accretion onto the black hole from the Bondi sphere I is continuing, then the matter will pile up
between the region of $R_{\rm Bondi}^{\rm II}\le R\le R_{\rm Bondi}^{\rm I}$. Such a dense shell will collapse 
when pressures are not powerful enough to balance it. Thus the accretion onto the seed black holes 
could be pulsive with a period. However, 
the oscillating accretion will be avoided if the outflow is supplied.

\subsection{Feedback: outflow}
Wind/outflow is a generic property of the accretion disk. There are many pieces of evidence for the presence 
of strong outflows from accretion disks. In the quasar PG 1211+143, the outflow is indicated by 
the blueshifted X-ray absorption lines with a velocity of $0.1c$ and has an outflow mass rate 
$\dot{M}_{\rm out}\sim 1.6\sunm/{\rm yr}$  (Pounds et al. 2003). Similar strong
outflows are found in ultraluminous X-ray sources (Mukai et al. 2003, Fabbiano et al. 2003)
and SS 433 which clearly has strong outflows ($>5\times 10^{-7}\sunm{\rm yr^{-1}}$) from the highly 
super-critical accretion disk (Kotani et al. 1996). A strong outflow is expected to develop from the tiny 
accretion disk, leading to efficient kinetic feedback to its 
surroundings. Here we neglect the detailed micro-physics of the interaction between outflow and medium. The 
kinetic luminosity is given by $L_{\rm kin}\approx (\dot{m}f_m\beta^2)L_{\rm Edd}\approx L_{\rm Edd}$, where 
$f_m$ is the fraction of the outflow to the accretion rate and we take $\dot{m}f_m\beta^2=1$. The simple 
energy budget can provide a rough estimation of the outflow kinetic energy. When the 
outflow is damped by its surroundings, its most kinetic energy will dissipate at the sonic radius
\begin{equation}
R_{\rm Sonic}\approx 1.0 {\rm pc}~L_{40}^{1/2}(n_4/6)^{-1/2}T_{0.1}^{-1/4},
\end{equation}
where $T_{0.1}=T/0.1{\rm keV}$ is the temperature of the surroundings (Begelman 2004). This radius is much 
smaller than the characterized radius of the fat disk ($R_d\sim 0.2{\rm kpc}$), but it is comparable to the 
inner quasi-spheric part of the cold fat disk. The kinetic
luminosity ($L_{\rm kin}$) can compensate the free-free radiation loss of the hot plasma within the radius 
$R_{\rm Sonic}$. For a typical outflow, the timescale of reaching the sonic point is  
$R_{\rm Sonic}/\beta c\approx 30(\beta/0.1)$yr. It would be interesting to note that this feedback timescale  
is much shorter than the Compton heating in the present case. We could draw a conclusion that 
the kinetic feedback from the outflow does play an important role in the growth of the seed hole.

On the other hand, the mass loss through the strong outflow definitely prevents 
the seed holes from growth. However the mass rate of the outflow is uncertain, we do not know 
exactly how the growth of the seed holes is affected via mass loss.

Finally, we would like to point out that we use the self-similar solutions of the super-Eddington accretion 
(Wang \& Zhou 1999) based on
the slim disk (Abramowicz et al 1988). The super-critical accretion remains a matter of debate since there are
several instabilities, especially the photon bubble instability (Gammie 1998). The instability results
in inhomogeneity of the disk, allowing the trapped photons to escape from the disk and enhance the
disk radiation. If we employ the radiation leakage model at a level of $L_{\rm Bol}\sim 300 L_{\rm Edd}$ 
(Begelman 2002), the feedback will be much stronger than we have discussed
above. Thus the results from the present paper is a lower limit of feedback. 

\section{Discussions}
For simplicity, we assume the radiation is isotropic from the tiny accretion disk. The anisotropic radiation 
from the slim disk could suppress the Compton heating in some degrees. The self-similar 
solution gives the ratio of the height to the radius $H/R=(5+\alpha^2/2)^{-1/2}\approx 0.45$, where $\alpha$ is 
the viscosity parameter (Wang \& Zhou 1999). The half-opening angle is about $\theta\approx 66^{\circ}$, the 
radiation will be confined within $1-\cos\theta\approx 60\%$ of the $4\pi$ solid angle. This assumption is 
reasonable. However the anisotropic radiation from the tiny disk may lower the feedback of the Compton heating.
Future work will consider how the anisotropy effects the feedback for a time-dependent
situation, which is beyond the scope of the present paper. 
 
There are other possible ways to form supermassive black holes. The direct collapse of primordial clouds in 
which the angular momentum is removed by the cosmic background photons (Loeb 1993, Loeb 
\& Rassio 1994) can form a $10^6\sunm$ black hole at $z\sim 15$. If the seed black hole accrets dark matter,
its growth is not suffering from the strong feedback (Hu et al. 2005). The inevitable fate 
of a compact cluster is the formation of a supermassive black hole (Quinlan \& Shapiro 1990). These possible 
ways might apply to different environments in the universe, it depends on future observations to test them.
 
\section{Conclusions}
We show that the feedback of the radiation pressure, Compton heating and outflows from the super-critical 
accretion disks will result in strong influence on the accretion itself. We show that the Compton heating 
almost quenches the super-critical accretion and the outflow from the tiny disk heats up the outer region 
so that accumulation of matter is avoided.  A rapid growth of seed black holes is greatly suppressed 
and they are hardly able to grow up unless the strong feedback can be avoided, for example considering the 
anisotropy of the radiation from the tiny accretion disk.

\acknowledgements{Useful comments from M. Volonteri 
are acknowledged. We are grateful to an anonymous referee for the helpful report. J. M. W\@. thanks the 
supports from a Grant for Distinguished Young Scientist from NSFC, NSFC-10233030.}

\newpage

\end{document}